\documentclass[twocolumn,english]{llncs}
\usepackage[T1]{fontenc}
\usepackage[latin9]{inputenc}
\usepackage{geometry}
\usepackage{graphicx}

\makeatletter
\newenvironment{lyxlist}[1]
{\begin{list}{}
{\settowidth{\labelwidth}{#1}
 \setlength{\leftmargin}{\labelwidth}
 \addtolength{\leftmargin}{\labelsep}
 }}
{\end{list}}

\usepackage{babel}
\makeatother

\begin{document}

\title{Meta-Packages: Painless Domain Specific Languages}
\titlerunning{Meta-Packages}

\author{Tony Clark }
\institute{Department of Computer Science, Middlesex University, London, UK\\
  \email{t.n.clark@mdx.ac.uk}
}
\maketitle
\begin{abstract}
Domain Specific Languages are used to provide a tailored modelling
notation for a specific application domain. There are currently two
main approaches to DSLs: standard notations that are tailored by adding
simple properties; new notations that are designed from scratch. There
are problems with both of these approaches which can be addressed
by providing access to a small meta-language based on packages and
classes. A meta-modelling approach based on \emph{meta-packages} allows
a wide range of DSLs to be defined in a standard way. The DSLs can
be processed using standard object-based extension at the meta-level
and existing tooling can easily be defined to adapt to the new languages.
This paper introduces the concept of meta-packages and provides a
simple example.
\end{abstract}

\section{Introduction}

The aim of any notation for software system design must be to faithfully
capture the structure and behaviour of the proposed system. The aim
of any tooling that supports such a notation must be that it is reliable
and provides useful functionality. What kind of functionality? Certainly,
the notation must be a good vehicle for communication - other people
must be able to understand your ideas expressed in the notation. However,
communication is not sufficient; the notation must support other useful
tasks that contribute to the development of a system. A design notation
must allow a tool to provide feedback on whether the system is likely
to work, perhaps even allow a tool to animate parts of the design.
In addition, it should be possible to generate parts of a system from
its design.

In recent years there has been interest in Model Driven Architecture
\cite{MDA}, Software Factories \cite{SWF}, tool definition languages
and Domain Specific Languages. All of these initiatives aim to provide
design notations that achieve the aims outlined above. These technologies
fall broadly into two categories: standard vs. extensible languages.

\subsection{Standard Modelling Notations}

The design notation is standard. UML 2.X is an example of this category
whereby the notation is standard with some limited scope for extension
points. In the case of UML there is a very large number of different
types of design element. Each type of element can be stereotyped by
defining properties and changing the iconization. This is similar
to \emph{annotations} recently added to Java. The advantages of this
approach are that the design notation is completely standard in terms
of how it is presented to the user, therefore tools mature quickly,
the expertise required to use the technology is relatively low, and
information is interoperable between tools. The disadvantage is that
the notation designers must preempt all the element types that will
be required, leading to notational bloat, and the scope for notational
extension is very limited.

\subsection{User Extensible Notations}

The design notation is user-defined. There are a number of technologies
including Microsoft Visual Studio DSL tools \cite{VSDSL}, GMF \cite{GMF}
and MetaEdit+ \cite{MetaEdit+} that allow a user to define new notation.
The extent to which the notation can be extended differs between technologies:
some tooling limits the definition through wizards and some allows
arbitrary extension through program-level interfaces. In most cases,
the underlying model for the design notation (the so-called meta-model)
is defined by the user and therefore the data representation is tailored
to the application domain. An advantage of this approach is that both
the notation and underlying data representation are a good fit for
the application domain, therefore tooling can take advantage of this
by supporting the use of the notation. The main disadvantages of this
approach are that the skill levels required to work with technology
are relatively high and that the graphical tooling must be developed
(even model-based) for each new language.

\subsection{The Best of Both Worlds}

Is it possible to achieve the advantages of each approach without
the disadvantages? Analysis of the use of many design notations leads
to the conclusion that many features recur while others differ. In
most cases there are package-like and class-like elements with association-like
relationships between them. Variations occur due to different categories
of these basic features with their own specific properties. 

This paper describes an approach to domain specific languages that
allows arbitrary extension of a small collection of underlying modelling
concepts via access to a self-describing meta-model. It shows that
by adding the notion of \emph{meta-packages} to the meta-language,
tooling can be designed that offers domain specific languages without
wholesale definition of a new language (such as that required by EMF
and GMF). The need for DSL tooling is discussed further by Fowler
\cite{DSLWB}.The result is an extensible modelling notation that
reuses the same tooling. The use of meta-package based technology
does not require skills in complete language definition and is not
limited to the simple property-based extensions of the standards based
approach described above. In addition, by extending the notion of
a package slightly, many new language features can be added.

Meta-packages are implemented in XMF-Mosaic and have been used on
a number of projects including generating code for telecomms applications
\cite{OSS,NGOSS,Integration,OSS2,Policy}. The screen-shots and code
in this paper are taken from a tutorial example that is implemented
in XMF-Mosaic. The book \cite{MetaModelling} provides a good introduction
to meta-modelling concepts.

The rest of this paper is structured as follows: section \ref{sec:Example}
provides the motivation for meta-packages using a simple example;
section \ref{sec:MetaPackages} defines meta-packages and how associated
tooling accommodates them; section \ref{sec:BeanDSL} implements the
example DSL using meta-packages; finally, section \ref{sec:Review}
reviews meta-packages, describes some extensions and discusses related
systems.

\section{Example Application}

\label{sec:Example}

\begin{figure}
\hfill{}\includegraphics[scale=0.55]{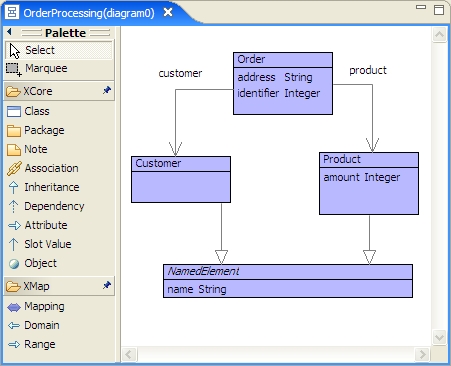}\hfill{}

\caption{Order Processing}
\label{fig:OrderProcessing1}
\end{figure}

Design notations should support the generation of source code where
this is sensible. The generation of source code from standard modelling
notations raises a problem. One of the merits of a modelling notation
is that it abstracts away from implementation details; however, implementation
details are required in order to generate the code. In effect, an
arbitrary number of sub-categories of modelling element needs to be
identified and represented in the design. Each category can be defined
in terms of properties, structure or behaviour.

Standard modelling notations often allow categories to be defined
using simple properties. However, this is not addressing the real
issue which is to be able to define new categories of modelling element
with their own structure and behaviour. Meta-packages support the
definition of new element categories by allowing a standard modelling
language to be extended at the meta-level. This section motivates
the definition of meta-packages using a simple example.

Java has recently been extended with annotations which can be used
to add properties to standard Java program elements such as classes.
The motivation for annotations has been the need to \emph{mark-up}
Java components with static information that can be used by tooling.
An example use of annotations is in the implementation of enterprise
information systems where Java class use annotations to define a mapping
to relational database tables. For example the simple model defined
in figure \ref{fig:OrderProcessing1} may give rise the to following
Java code:

\begin{verbatim}
@Entity
@Table(name="ORDER_TABLE")
public class Order {

  private int id;
  private String address;

  @Id
  @Column(name="ORDER_ID")
  public int getId() {
    return id;
  }
  public void setId(int id) {
    this.id = id;
  }

  @Column(name="SHIPPING_ADDRESS")
  public String getAddress() {
    return address;
  }
  public void setAddress(String address) {
    this.address = address;
  }
}
\end{verbatim}

\noindent in which instances of the class Order are to be persisted
in a relational database as rows in the table named ORDER\_TABLE with
columns ORDER\_ID and SHIPPING\_ADDRESS. The primary key of this table
is ORDER\_ID.

\begin{figure}
\hfill{}\includegraphics[scale=0.45]{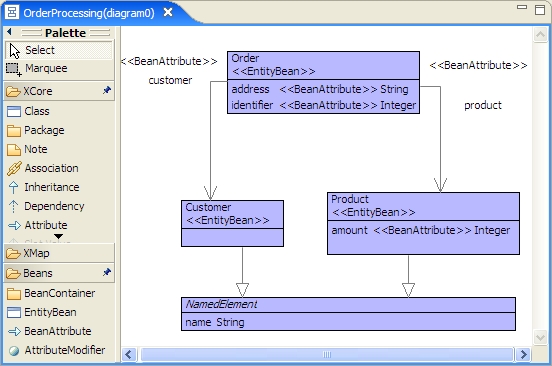}\hfill{}

\caption{Bean Modelling Tool}
\label{fig:BeanModelling}
\end{figure}

The tool used to represent the model in figure \ref{fig:OrderProcessing1}
is standard since it provides tools in the palette to create basic
modelling elements: Package, Class, Attribute etc. Given a model using
this tool, there is no way to distinguish between those model element
that will become entity beans and those that will become basic Java
classes.

Figure \ref{fig:BeanModelling} shows a tool that provides a new category
of modelling element: Beans. The model for the order processing system
now includes two different categories of modelling element: Classes
and Attributes, EntityBeans and BeanAttributes. It is now possible
to distinguish between the elements in terms of code generation. In
addition, EntityBean will have a property that defines the name of
the relational database table used to represent its instances.

\begin{figure}
\hfill{}\includegraphics[scale=0.5]{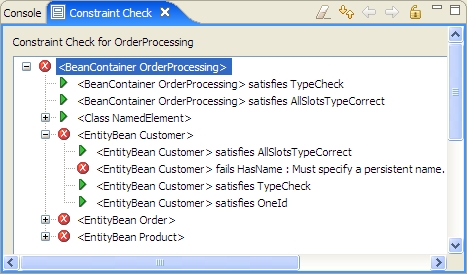}\hfill{}

\caption{DSL Semantics}
\label{fig:DSLSemantics}
\end{figure}

In addition to defining new modelling entities, a DSL has semantics.
Part of the semantics of a language are the rules that govern correct
model formation. These are expressed at the meta-package level, for
example an EntityBean is correctly formed when it has at most one
bean attribute that is designated as an id (the primary key in the
table). Figure \ref{fig:DSLSemantics} shows the result of running
the well-formedness checks for the meta-package over the instance
shown in figure \ref{fig:BeanModelling}. Leaves of the tree are constraints
that have been applied to the model elements. Triangles represent
constraints that are satisfied. Crosses are constraints that have
failed. Therefore, in the example model the Customer element has OneId
(a single attribute designated as a primary key) but does not specify
a persistent name (a table in the database).

Meta-packages provide a mechanism for extending the basic language
for class-based modelling shown in figure \ref{fig:OrderProcessing1}
with \emph{domain specific} features such as those shown in figure
\ref{fig:BeanModelling}. Tooling for standard modelling detects meta-package
automatically and extends the functionality appropriately.

\section{Meta-Packages}

\label{sec:MetaPackages}

Meta-packages are a way of defining semantically rich DSLs without
having to specify associated tooling. The idea is that there is a
single basic meta-package that defines a language. The concepts in
the base meta-package can be extended to produce new meta-packages.
Each meta-package is a language and tooling is written against the
base meta-package. Since all new meta-packages are extensions of the
base, the existing tooling will work with any new language. Since
languages are written at the meta-level, executable meta-modelling
allows any new language to have a rich semantics.

This section defines meta-packages and is structured as follows: section
\ref{sub:XCore} defines the base meta-package called XCore; section
\ref{sub:Tools} defines a tool model for languages over XCore; section
\ref{sub:Diagrams} defines a model for tool diagrams; section \ref{sub:Displays}
defines a model for the display elements on diagrams; finally, section
\ref{sub:Mappings} defines mappings that ensure the diagrams and
model elements are synchronized in the tools.

\subsection{XCore}

\label{sub:XCore}

\begin{figure}
\hfill{}\includegraphics[scale=0.4]{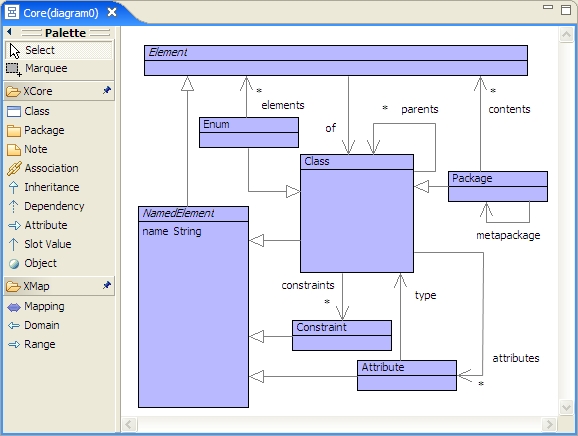}\hfill{}

\caption{XCore}
\label{fig:XCore}

\end{figure}

Meta-packages are defined in a meta-circular language defined in figure
\ref{fig:XCore}. This language is part of the basic package, called
XCore, of the XMF-Mosaic modelling tool. In XMF-Mosaic, everything
is an element. Classes are elements with names, constraints, attributes
and parents. Note that every element has a type (\emph{of}) which
is a class. A Package is a class that contains elements and which
has a meta-package. The idea of the meta-package association is that
each element contained by a package is \emph{of} some class contained
by its meta-package. By default, the meta-package of all packages
is XCore.

\noindent The semantics of XCore is defined using a collection of
constraints. Since XCore is self defining these constraints are defines
with respect to XCore itself. The following is an example which states
that all elements of an enumerated type must be instances of that
type and no other instances exist:

\begin{verbatim}
context Enum
  elements->forAll(e | e.of = self)
context Element
  of.of = Enum implies
    of.elements->includes(self)
\end{verbatim}

\noindent XCore supports executable meta-modelling via XMF. The following
operations are used in the rest of this paper:

\begin{verbatim}
context Class
  @Operation allParents():Set(Class)
    parents->iterate(p P = Set{} | 
      P + p.allParents())
  end
context Package 
  @Operation modellingElements():Set(Class)
    elements->select(e |
      e.isKindOf(Class))
  end
context Element
  @Operation tag(expected:Class):String
    if of = expected
    then ""
    else of.name
    end
  end
\end{verbatim}

\subsection{Tools}

\label{sub:Tools}

\begin{figure}
\hfill{}\includegraphics[scale=0.5]{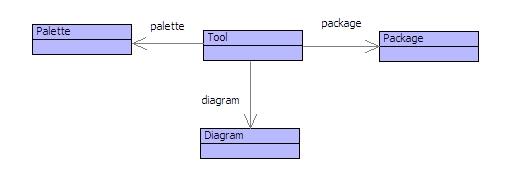}\hfill{}

\caption{Tool Models}
\label{fig:ToolModels}
\end{figure}

Modelling in XCore-based languages is supported by tooling. Figure
\ref{fig:XCore} shows an example of a modelling tool that consists
of a palette on the left-hand side and a diagram containing modelling
elements on the right. The palette contains buttons for the modelling
elements defined in the modelling language. The tool palette depends
on the modelling language's meta-package; however there is only one
tool engine which is parameterized with respect to the meta-package. 

Figure \ref{fig:ToolModels} shows the key elements involved in defining
a modelling tool. A tool associates a package of modelling elements,
a diagram and a palette of button groups. Buttons are used to select
creation modes. Events on the diagram modify diagram elements and
events to the package modify modelling elements. Daemons on both the
diagram and the package ensure that events from the model are propagated
to the diagram and vice versa.

\begin{figure}
\hfill{}\includegraphics[scale=0.5]{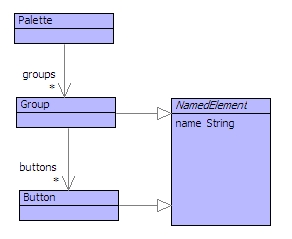}\hfill{}

\caption{Tool Palette}
\label{fig:ToolPalette}

\end{figure}

Figure \ref{fig:ToolPalette} shows the structure of a tool palette.
The palette consists of groups of buttons which are selected in order
to determine the creation mode on a diagram. Each group corresponds
to a meta-package. All packages have XCore as a meta-package by default
and therefore have the XCore and XMap groups. If the meta-package
of the package associated with a tool is P which inherits from XCore
then the palette will have groups named XCore, XMap and P. This leads
to the following constraint:

\begin{verbatim}
context Tool
  palette.groups.name = 
  package.metaPackage.allParents().name
\end{verbatim}

\noindent The buttons provided by each group are determined by the
language elements defined in the meta-package. A modelling element
is a sub-class of Package, Class or Attribute as defined in XCore.
Therefore:

\begin{verbatim}
context Tool
  package.metaPackage.allParents()
    ->forAll(p |
      palette.groups
        ->exists(g |
           g.name = p.name and
             p.modellingElements().name =
               g.buttons.name))
\end{verbatim}

\subsection{Diagrams}

\label{sub:Diagrams}

\begin{figure}
\hfill{}\includegraphics[scale=0.5]{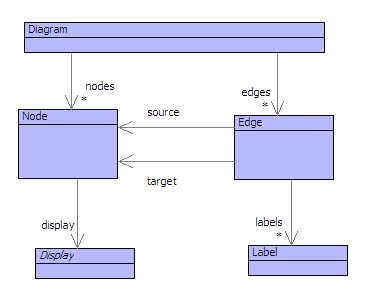}\hfill{}

\caption{Diagram Models}

\label{fig:DiagramModels}
\end{figure}

Each tool manages a diagram that consists of nodes and edges. Each
node has a display that is used to render the node on the diagram.
Each edge has a source and a target node and a collection of labels.
The diagram model is shown in figure \ref{fig:DiagramModels}. The
tool requires that the diagram is always synchronized with the package.
Therefore the following constraint must always hold:

\begin{verbatim}
context Tool
  package.classes()->size =
  diagram.nodes()->size
\end{verbatim}

\noindent The constraint on edges is less easy to define since attributes
may be shown on a diagram within classes or as edges. In addition,
inheritance is shown as edges. To make matters more complex the types
of attributes and the target of inheritance edges may be imported
from other packages.

\subsection{Displays}

\label{sub:Displays}

\begin{figure}
\hfill{}\includegraphics[scale=0.5]{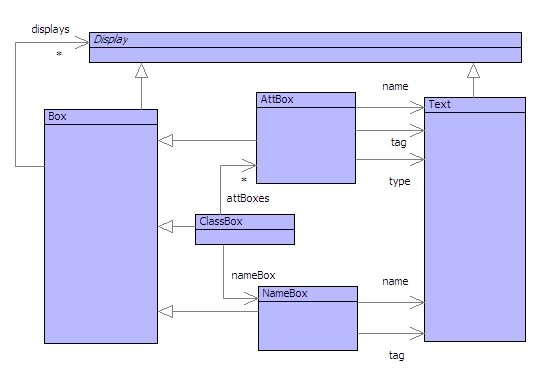}\hfill{}

\caption{Displays}
\label{fig:Displays}
\end{figure}

Figure \ref{fig:Displays} shows some of the display types that can
be used to render nodes on diagrams. Two basic types of display are
shown: boxes and text. Package-based diagrams use specializations
of these display types to render classes. A class box consists of
a name box and several attribute boxes. Note that attBoxes and nameBox
are derived attributes since they are just specifically identified
components of the displays. Both attrbute boxes and name boxes have
some text for the name of the element. They also both have a tag which
is used to identify the meta-type of the element if it not Class or
Attribute respectively. An attribute box has an additional text field
for the type of the attribute.

\subsection{Mappings}

\label{sub:Mappings}

\begin{figure*}
\hfill{}\includegraphics[scale=0.7]{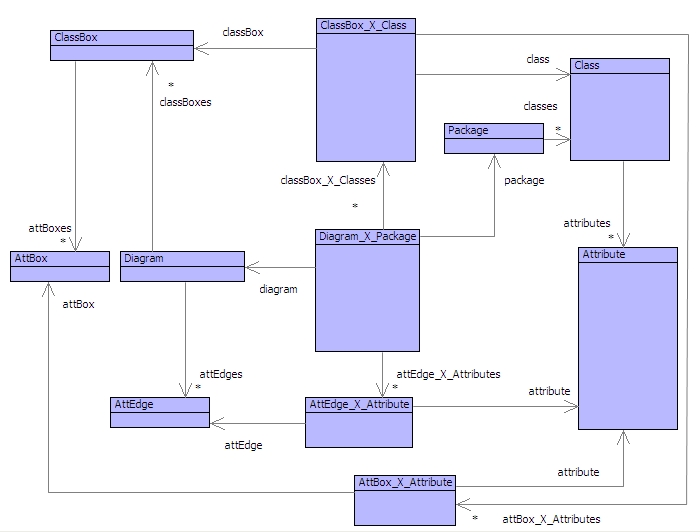}\hfill{}

\caption{Mapping}
\label{fig:Mapping}
\end{figure*}

Diagram-model synchronization is handled by a collection of mappings
associated with a tool. Figure \ref{fig:Mapping} shows the mapping
model for package diagrams. A mapping has the form A\_X\_B which associates
instances of A with instances of B. Each mapping class has a constraint
that defines the synchronization requirements. Many of these constraints
state commutative properties of the model. For example:

\begin{verbatim}
context Tool
  mapping.package = package and
  mapping.diagram = diagram
\end{verbatim}

\noindent Each of the mappings inherit from a class OneToOne (not
shown) which requires that all maplets are unique. A mapping instance
can reference all of the elements in the domain and range of the mapping
and also reference all of the associations (one of which is itself).
The following constraints require the OneToOne mapping to have unique
associations:

\begin{verbatim}
context OneToOne
  domain->forAll(d |
    range->exists(r1 r2 |
      map(d,r1) and map(d,r2) 
        implies r1 = r2))

context OneToOne
  range->forAll(r |
    domain->exists(d1 d2 |
      map(d1,r) and map(d,r2)
        implies d1 = d2))
\end{verbatim}

\noindent All of the classes in the package must be shown on the diagram
and all class nodes on the diagram must correspond to some class in
the package:

\begin{verbatim}
context Diagram_X_Package
  package.classes = classBox_X_Classes.class
context Diagram_X_Package
  diagram.classBoxes = 
  classBox_X_Classes.classBox
\end{verbatim}

\noindent All the attribute edges must be related and must correspond
to an attribute. Note that not all attributes need to be shown as
edges on the diagram:

\begin{verbatim}
context Diagram_X_Package
  attEdge_X_Attributes.attEdge = 
  diagram.attEdges
context Diagram_X_Package
  attEdge_X_Attributes.attribute
    ->subSet(package.classes.attributes)
\end{verbatim}

\noindent The name of a class on a diagram must always be synchronized
with the name of the corresponding model element. If the element is
directly an instance of class then its tag is empty otherwise the
tag must be the name of the meta-type:

\begin{verbatim}
context ClassBox_X_Class
  classBox.name = class.name
context ClassBox_X_Class
  classBox.tag = class.tag
\end{verbatim}

\noindent All of the attributes in a class box must correspond to
some attribute of the associated class. However, the not all attributes
need to be displayed in a class box:

\begin{verbatim}
context ClassBox_X_Class
  attBox_X_Attributes.attribute
  ->subSet(class.attributes)
context ClassBox_X_Class
  attBox_X_Attributes.attBox =
  classBox.attBoxes
\end{verbatim}

\noindent Attribute boxes offer a name, a type and an optional meta-tag.
The following constraints require that these are always synchronized
(attribute edge tags are specified using similar constraints):

\begin{verbatim}
context AttBox_X_Attribute
  attribute.name = attBox.name
context AttBox_X_Attribute
  attribute.type.name = attBox.type
context AttBox_X_Attribute
  attribute.tag = attBox.tag
\end{verbatim}

\noindent Finally, any attribute edges that are shown on the diagram
must correspond to attributes. The source of the edge must be a class
box that corresponds to the owner of the attribute and the target
of the edge must be a class that corresponds to the type of the attribute.
The predicate isClass is not defined but is satisfied isClass(b,c)
when the box b is associated with the class c in the tool:

\begin{verbatim}
context AttEdge_X_Attribute
  isClass(attEdge.sourceBox,attribute.owner)
context AttEdge_X_Attribute
  isClass(attEdge.targetBox,attribute.type)
\end{verbatim}

\noindent All of the attributes must be shown as either edges or in
boxes:

\begin{verbatim}
context Diagram_X_Package
  package.classes.attributes =
    attEdge_X_Attributes.attribute +
    classBox_X_Classes.attBox_X_Attributes.attribute
\end{verbatim}

\noindent Other constraints can be defined to require that an attribute
edge cannot be shown as a boxed attribute.

\section{A Bean DSL}

\label{sec:BeanDSL}

The previous section has defined the meta-package relationship that
allows new class-based DSLs to be defined by extending meta-concepts.
Existing tooling mechanisms can accommodate the new DSLs without any
new code being necessary. This section shows how a new language to
support Beans for Enterprise Systems can be defined using the meta-package
relationship. The result of the definition is a new tool as shown
in figure \ref{fig:BeanModelling}.

This section is defined as follows: section \ref{sub:AbstractSyntax}
describes the abstract syntax for the DSL; section \ref{sub:ConcreteSyntax}
describes how concrete syntax can be defined in a number of ways;
section \ref{sub:ConstraintChecking} describes well-formedness semantics
for the DSL; section \ref{sub:CodeGeneration} defines how Java code
is generated from the DSL.

\subsection{Abstract Syntax}

\label{sub:AbstractSyntax}

\begin{figure*}
\hfill{}\includegraphics[scale=0.6]{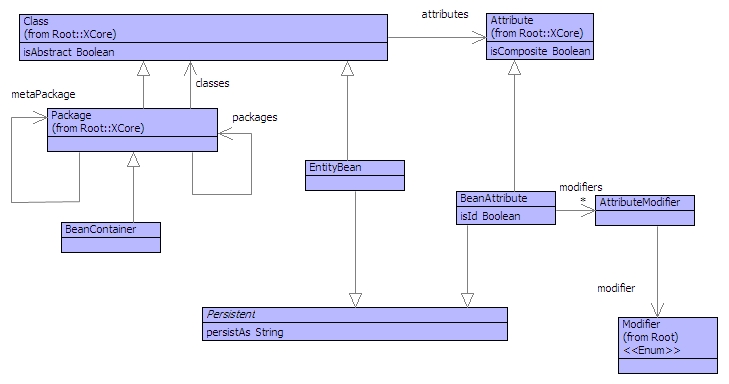}\hfill{}

\caption{Beans}

\label{fig:Beans}
\end{figure*}

The first step in DSL definition is to define a new meta-package whose
modelling concepts extend those of an existing meta-package. Figure
\ref{fig:Beans} shows the definition of the package named Beans.
The concepts are defined as follows:

\begin{lyxlist}{00.00.0000}
\item [{\textbf{BeanContainer}}] A bean container is a specialization of
Package. The contents of a bean container may be entity beans and
the class BeanContainer provides us with a container for bean specific
constraints.
\item [{\textbf{EntityBean}}] An entity bean is a class that has a persistsAs
property that is used to name the relational database table used to
contain all the instances of the class.
\item [{\textbf{BeanAttribute}}] A bean attribute is a special type of
attribute that names the column in the database table used to contain
its values. A bean attribute can be tagged as being a primary key
in the relational table by setting its isId attribute. A bean attribute
may also require a Java accessor and updater. These properties are
set via the attribute modifiers of the bean attribute.
\end{lyxlist}
\begin{figure}
\hfill{}\includegraphics[scale=0.6]{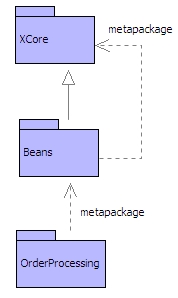}\hfill{}

\caption{Declaring Beans a Meta-Package}

\label{fig:MetaPackage2}
\end{figure}

The Beans package must be designated as a meta-package. This is done
by making it \emph{inherit from} XCore as shown in figure \ref{fig:MetaPackage2}.
By default, a package inherits from an empty package and is therefore
not a meta-package. By making a package inherits from a meta-package,
the new package also becomes a meta-package.

The package OrderProcessing is to be written in the language defined
by Beans. By default the meta-package of a new package is XCore. Figure
\ref{fig:MetaPackage2} shows the meta-package of OrderProcessing
has been changed to Beans. Note that the super-package of OrderProcessing
has not been changed, so it stays as the default empty package and
therefore OrderProcessing is not itself a meta-package.

\subsection{Concrete Syntax}

\label{sub:ConcreteSyntax}

A DSL does not require a concrete syntax, but it is generally important
for usability to have one. DSLs that are defined via meta-packages
get a concrete diagram syntax for free. The tooling that is specified
in section \ref{sec:MetaPackages} detects the new modelling concepts
and provides new palette buttons and labelling as appropriate.

In addition to diagram syntax for a language, textual syntax is often
useful. XMF \cite{Superlanguages} has a text language for package
definition that supports the declaration of meta-package information.
This is exactly the same as the diagram tooling: the same text processing
engine is used even though the language has changed via meta-package
extension. The following shows how the OrderProcessing bean container
can be defined using XMF package definition:

\begin{verbatim}
@Package OrderProcessing 
  metaclass BeanContainer metapackage Beans 
  @Class NamedElement isabstract
    @Attribute name : String end
  end
  @Class Order metaclass EntityBean
    @Attribute identifier 
      metaclass BeanAttribute : Integer 
    end
    @Attribute address 
      metaclass BeanAttribute : String 
    end
    @Attribute customer 
      metaclass BeanAttribute : Customer 
    end
    @Attribute product 
      metaclass BeanAttribute : Product 
    end
  end
  @Class Customer metaclass EntityBean 
    extends NamedElement
  end
  @Class Product metaclass EntityBean 
    extends NamedElement
    @Attribute amount 
      metaclass BeanAttribute : Integer 
    end
  end
end
\end{verbatim}

\noindent A problem with the basic text definition shown above is
that it does not provide support for setting the new meta-properties
such as persistAs. These can be set independently, but it is much
better if an entity can be defined as a modular unit. In addition,
the above syntax exposes implementation details about the modelling
concepts by requiring their meta-classes to be specified.

XMF supports extensible syntax via syntax-classes \cite{Superlanguages}
which are normal classes that define grammars for processing text.
This mechanism can be used to define a new concrete syntax for a DSL
that translates into the basic definitions given above. The following
is a DSL for beans that has been used to define the order processing
system:

\begin{verbatim}
@BeanContainer OrderProcessing
  entity NamedElement
    name       (NAME)             : String
  end
  entity Order(ORDER_TABLE) [NamedElement]
    *identifier(ORDER_ID)         : Integer
    address    (SHIPPING_ADDRESS) : String
    customer   (CUSTOMER_REF)     : Customer
    product    (PRODUCT_REF)      : Product
  end
  entity Customer extends NamedElement
  entity Product extends NamedElement
    amount     (AMOUNT)           : Integer
  end
end
\end{verbatim}

\noindent The example above hides all of the implementation detail.
This is achieved by defining a new syntax class for BeanContainer
as shown below:

\begin{verbatim}
@Class BeanContainer
  @Grammar
    BeanContainer ::= n = Name es = Entity* {
     [| let P = @Package <n> 
                  metaclass BeanContainer 
                  metapackage Beans end
        in <es->iterate(e x = [| P |] | 
             [| P.add(<e>); <x> |])>
        end |]
    }.
    Entity ::= 
     n = Name p = Persist s = Super as = Att* {
     [| let C = @Class <n> 
                  metaclass EntityBean
                  extends <s>
                end
        in C.persistAs := <p.lift()>;
              <as->iterate(a x = [| C |] | 
                [| <x>.add(<a>) |])>
           end
        |]
    }.
    Persist ::= '(' Name ')'.
    Super ::= '[' Type ']'.
    Att ::= 
      i = IsId n = Name p = Persist ':' t = Type {
      [| let A = BeanAttribute(<n.lift()>,<t>)
         in A.persistAs := <p.lift()>
         end |]
    }.
  end
end
\end{verbatim}

\subsection{Constraint Checking}

\label{sub:ConstraintChecking}

Constraint checking involves executable modelling (being able to attach
executable predicate expressions to classes as shown in figure \ref{fig:XCore}).
Elements are well-formed when they satisfy all of the constraints
defined by their class. In addition, containers, such as packages,
are well formed when their contents are well-formed.

In order for a bean container to be well-formed, all of the persistent
elements must specify a name that can be used in the relational database:

\begin{verbatim}
context Persistent 
  @Constraint HasName
    persistAs <> ""
    fail "Must specify a persistent name."
  end 
\end{verbatim}

\noindent An entity bean is well formed when there is at most one
bean attribute that designates a primary key:

\begin{verbatim}
context EntityBean 
  @Constraint OneId
    not @Exists a1,a2 in attributes ->
          a1 <> a2 and 
          a1.isId and 
          a2.isId
        end
    fail "Cannot have multiple ids."
  end 
\end{verbatim}

\subsection{Code Generation}

\label{sub:CodeGeneration}

Code generation involves a mapping from a model to source code. This
is the essence of MDA in which UML models are translated to code.
However, UML does not allow access to the meta-level and therefore
the scope for extending the modelling language with sophisticated
translation mechanisms is limited.

To translate from elements in figure \ref{fig:Beans} to the source
code shown in section \ref{sec:Example} we can also use executable
meta-modelling technology. A code export operation is defined for
each class that is to be translated. XMF provides \emph{code-template}
technology that makes it easy to generate code \cite{Superlanguages}.
The following template:

\begin{verbatim}
@Java(out,leftMargin)
  class <n> { 
  }
end
\end{verbatim}

\noindent writes an empty class definition to the output channel,
newlines tab to leftMargin. Within the Java code template all text
is faithfully written to the output except for expressions delimited
by < and >. Such expressions are evaluated and then written to the
output channel. In the example above, the generated class will have
a name that is the value of the variable n. Within expressions, the
use of {[} and ] delimiters switch back to literal code and nesting
of {[}, ] and <,> is permitted.

Entity beans are translated to code as follows:

\begin{verbatim}
context EntityBean
  @Operation code(out:OutputChannel)
    @Java(out,7)
      @Entity
      @Table(name="<persistAs>")
      public class <name>  {
        <@For a in attributes 
           when a.isKindOf(BeanAttribute) do
          [private <a.typeName()> <a.name>;]
         end;
         @For a in attributes 
           when a.isKindOf(BeanAttribute) do
          a.code(out)
         end>
      }
    end
  end
\end{verbatim}

\noindent Each bean attribute is translated to code as follows:

\begin{verbatim}
context BeanAttribute
  @Operation code(out:OutputChannel)
    let name = name.toString() then
        Name = name.upperCaseInitialLetter()
    in
      @Java(out,7)
        <if isId then [@Id] end>
        @Column(name="<persistAs>")
        <if self.canGet() 
         then self.getCode(out,name,Name) 
         end>
        <if self.canSet() 
         then self.setCode(out,name,Name) 
         end>
      end
    end
  end
    
context BeanAttribute
  @Operation getCode(out,name,Name)
    @Java(out,5)
      public <self.typeName()> get<Name>() {
        return <name>;
      }
    end
  end
    
context BeanAttribute
  @Operation setCode(out,name,Name)
    @Java(out,5)
       public void set<Name>(<self.typeName()> <name>) {
        this.<name>  = <name>;
      }
    end
  end
\end{verbatim}

\section{Review}

\label{sec:Review}

This paper has identified two broad approaches to DSLs for design
notations: standards based and user defined. Standards, notably the
UML family, offer mature tooling and interchangeable models, but can
be bloated and lack mechanisms for sophisticated extensibility. Technologies
for user defined notations offer arbitrary flexibility but at a cost
of complexity and starting from scratch each time.

Meta-packages is an meta-modelling based approach to defining languages
that allows tooling to be developed that does not require significant
modification each time a new language is developed. Since the approach
uses true meta-modelling, object-oriented techniques can be used to
define semantics for the new language features. Meta-packages have
been specified and a simple example DSL for Enterprise Systems has
been described. 

Meta-packages are an approach rather than a single technology. The
key features are a single meta-circular meta-model (XCore defined
here), executable modelling, and the meta-package relationship. This
combination of features guarantees that any tooling based on the base
meta-language will work with any new language that is defined. 

The XCore language as defined in this paper is rather small. Meta-packages
are supported by XMF-Mosaic where XCore is much larger, however the
principles are the same. One important feature supported by XMF-Mosaic
is the ability for diagrams to render element-nodes and slot-edges.
Since we advocate a true meta-modelling approach, everything is ultimately
an instance of the class Element. Packages can contain elements. If
diagrams can render elements and represent slot-values via edges then
\emph{any} package element can be represented on a diagram and related
to their owner. This feature guarantees that any language can be supported
by tooling that is parameterized with respect to the base language,
even if the element is not a specialization of the basic meta-concepts
(Package, Class, Attribute etc). For example, classes could be extended
to represent components with ports. A port can be represented on a
diagram as a basic element with a slot-value edge from the component
to the port. XMF-Mosaic is available open-source under EPL from the
Ceteva web-site (http://www.ceteva.com).

Virtually all other tools for DSL definition do not implement the
golden braid (meta-model-instance) \cite{GoldenBraid}. For example,
GMF models are instances of Ecore but cannot themselves have instances.
The same is true of UML and therefore UML tooling and of Visual Studio
DSL tools. The golden braid is a key feature in the meta-package approach
since it allows tooling to be defined that is reusable with models
that can be extended with their own semantics.

\end{document}